\documentstyle[times,graphics,astrobib,amssymb,epsfig]{mn2e}

\def\etal{et al.\ }
\def\ie{i.\,e.\,, }

\def\unit #1{\,{\rm #1}}

\def\arcmin{\unit{arcmin}}

\def\gne #1#2{\ \vphantom{S}^{\raise-0.5pt\hbox{$\scriptstyle#1$}}_
{\raise0.5pt \hbox{$\scriptstyle#2$}}}

\def\unit #1{\,{\rm #1}}

\title[Barred lenticulars: luminosity and environment dependence]
{Bar fraction in lenticular galaxies: dependence on luminosity and environment}

\author[Barway \etal]
{Sudhanshu Barway $^{1}$\thanks{E-mail: barway@saao.ac.za (SB)}, Yogesh Wadadekar $^{2}$\thanks{E-mail: yogesh@ncra.tifr.res.in (YW)}, and  Ajit K. Kembhavi $^{3}$\thanks{E-mail: akk@iucaa.ernet.in (AKK)}\\
$^{1}$South African Astronomical Observatory, P.O. Box 9, 7935, Observatory, Cape Town, South Africa; \\ 
$^{2}$National Centre for Radio Astrophysics, Post Bag 3, Ganeshkhind, Pune 411007, India; \\ 
$^{3}$Inter University Centre for Astronomy and Astrophysics, Post Bag 4, Ganeshkhind, Pune 411007, India. \\  } 

\begin{document}

\maketitle


\begin{abstract}

We present a study of bars in lenticular galaxies based on a sample of
371 galaxies from the SDSS-DR 7 and 2MASS in optical and near-infrared
bands, respectively. We found a bar in 15\% of the lenticular galaxies
in our sample, which is consistent with recent studies. The barred
galaxy fraction shows a luminosity dependence, with faint lenticular
galaxies ($M_K > -24.5$, total absolute magnitude in $K$ band) having a larger
fraction of bars than bright lenticular galaxies ($M_K < -24.5$). A
similar trend is seen when $M_r = -21.5$, the total absolute magnitude in SDSS
$r$ band is used to divide the sample into faint and bright lenticular
galaxies. We find that  faint galaxies in clusters show a higher bar
fraction than their counterparts in the field. This suggests that the  
formation of bars in lenticular galaxies not only depends on the total 
luminosity of galaxy but also on the environment of the host galaxy.
\end{abstract}


\begin{keywords}

galaxies: elliptical and lenticular -   fundamental parameters
galaxies: photometry - structure - bulges 
galaxies: formation - evolution
\end{keywords}


\section{Introduction}

The presence of bars has important implications for disk galaxies due
to their deep connection with the dynamical and secular evolution of
such galaxies (Kormendy \& Kennicutt 2004). N-body simulations and
many theoretical studies predict that bars transfer angular momentum
to the outer disk, which causes the stellar orbits in the bar to
become elongated and the bar amplitude to increase (Pfenniger \&
Friedli 1991; Sellwood \& Wilkinson 1993; Athanassoula 2003). The
growing bar becomes more and more efficient in driving gas in towards
the centre of the disk, which can trigger star-bursts (Hunt \& Malkan
1999; Sakamoto \etal 1999; Regan \& Teuben 2004) and contribute to the
formation of disky bulges or pseudo bulges (Kormendy \& Kennicutt
2004; Debattista \etal 2004; Athanassoula \etal 2005; Jogee \etal
2005; Sheth \etal 2005; Debattista \etal 2006). Bars are typically
dominated by evolved stellar populations (Gadotti \& de Souza 2006)
but sometimes they are also associated with enhanced nuclear and
circumnuclear star formation (Ho \etal 1997). Barred galaxies are
observed to have larger reservoirs of molecular gas in their centres
relative to unbarred galaxies (Sakamoto \etal 1999, Sheth \etal
2005). A recent study by Barazza \etal (2008), with a large sample,
found that the bar fraction is higher in blue, lower-luminosity,
late-type disks compared to more massive, red, early-type
galaxies. However, bars in early-type galaxies tend to be stronger,
more elongated and longer, both in an absolute sense and relative to
the size of the disk (Elmegreen \& Elmegreen 1985; Erwin \etal 2005;
Menendez-Delmestre \etal 2007). There have been a few studies on the
relation between the bar fraction and environment of the host galaxy
which suggest that the frequency of bar formation does not depend
significantly on host galaxy environment (Aguerri \etal 2009, Barazza
\etal 2009; Marinova \etal 2009; van den Bergh 2002). On the other
hand, some studies have suggested that barred galaxies are more
concentrated towards cluster centers than unbarred disks in rich
clusters like Coma and Virgo (Barazza \etal 2009; Thompson 1981;
Andersen 1996).

Optical and near-infrared imaging has revealed bars (both prominent
and weak) in a majority (50-70\%) of local disk galaxies,  including 
lenticular galaxies and irregulars, with a wide range of bulge-to-total 
luminosity ratio and mass (de Vaucouleurs 1963; Eskridge \etal 2000
Whyte \etal 2002; Marinova \& Jogee 2007; Menendez-Delmestre \etal
2007; Barazza \etal 2008). A substantial population of bars exists in 
lenticular galaxies (Nair \& Abraham 2010; Aguerri \etal 2009) which are 
not dynamically cool, indicating that the mechanism responsible for the 
formation of the bar is more complex than the accepted mechanism based 
on dynamical instabilities in cold stellar systems (Bournaud \& Combes 2002). 
The lenticular galaxies introduced by Hubble (1936) as a morphological 
transition class between elliptical and early-type spiral galaxies, which have 
most the massive bulges among disk galaxies, may have formed in several 
different ways as suggested by theoretical and numerical simulation studies 
(Bekki 1998; Abadi \etal 1999). 

Barway \etal (2009, 2007) have presented evidence to support the view
that the formation history of lenticular galaxies depends upon their
luminosity. According to this view, luminous lenticulars are likely to
have formed their bulges at early epochs through a rapid collapse
followed by rapid star formation, similar to the formation of
elliptical galaxies (Aguerri \etal 2005). On the other hand,
low-luminosity lenticular galaxies likely formed by the stripping of
gas from the disc of late-type spiral galaxies, which in turn formed
their pseudo bulges through secular evolution processes induced by
bars. If this is true there must be signatures of the formation mechanism 
imprinted in the light profile (correlated bulge disk sizes), stellar populations 
(as traced by colours), stellar kinematics (as traced by 3D spectroscopy) and
more evidently in the presence of a kinematic structure such as a
stellar bar.

\begin{figure}
\centerline{\psfig{figure=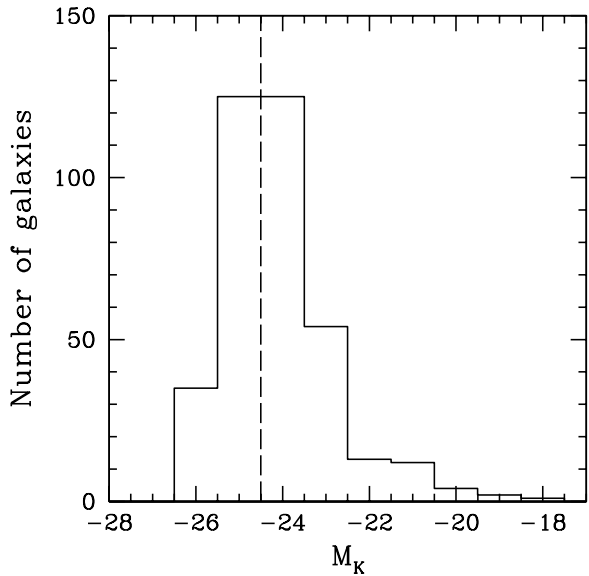,width=8.0cm,angle=0.}}
\caption{Distribution of total $K$ band absolute magnitude ($M_K$). 
The vertical dashed line corresponds to total absolute
  magnitude $M_K$ = -24.5, which we use to divide low- and
  high-luminosity lenticular galaxies in the near-IR.}
\label{f1}
\end{figure}


\begin{figure}
\centerline{\psfig{figure=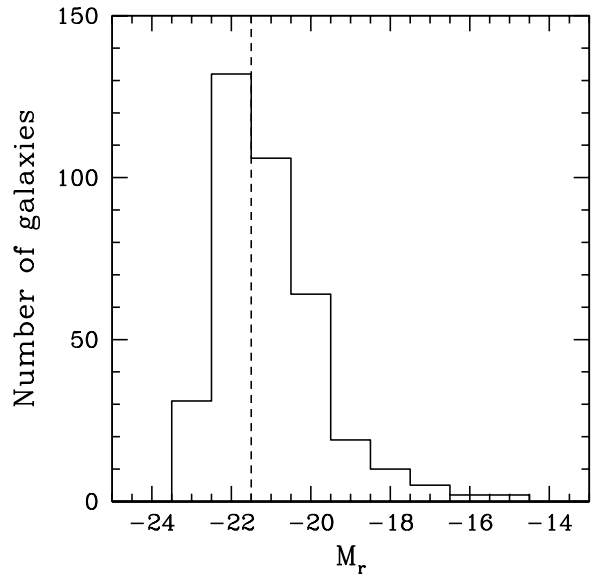,width=8.0cm,angle=0.}}
\caption{Distribution of total $r$ band absolute magnitude ($M_r$). 
The vertical dashed line corresponds to total absolute
  magnitude $M_r$ = -21.5, which we use to divide low- and
  high-luminosity lenticular galaxies in the optical.}
\label{f2}
\end{figure}


In this Letter, we present evidence for a significantly enhanced
probability of the existence of a bar in fainter lenticular galaxies
relative to brighter ones. We focus on the variation of the bar
fraction in lenticular galaxies with total luminosity, in $K$ band
as well as $r$ band, using 2MASS and SDSS data, respectively. We use a sample
of 371 lenticular galaxies in the local Universe in the present
study. Throughout this Letter, we use the standard concordance
cosmology with $\Omega_M = 0.3$, $\Omega_\Lambda = 0.7$, and $h_{100}
= 0.7$.

\section{The sample and data}

We aimed for a sample, from the field as well as cluster environments,
which is a fair representation of the lenticular (S0) galaxy
population in the near-by universe and has a statistically meaningful
number of galaxies spanning a large range of luminosities. We began by
selecting all galaxies with apparent blue magnitude brighter than $m_B
= 14$ and classified as lenticular in the Uppsala General Catalogue of
Galaxies (UGC; Nilson 1973). The UGC is essentially complete to a
limiting major-axis diameter of 1 $\arcmin$, or to a limiting apparent
magnitude of 14.5 on the blue prints of the Palomar Observatory Sky
Survey for the sky north of declination -2$^\circ$.5. This provides a
sample of 635 lenticular galaxies.  Next, we searched the Sloan
Digital Sky Survey Data Release 7 (SDSS-DR7; Abazajian \etal 2009) to
get data from that survey on our sample of lenticular galaxies. We
found that 387 lenticular galaxies have SDSS-DR7 imaging in five bands
($u$, $g$, $r$, $i$, $z$). All of these galaxies except two (which are
affected by artifacts in the 2MASS scans) have near-infrared data from
2MASS in $J$, $H$ and $K$ bands as well. We also use the {\it
  Hyperleda}\footnote{http://leda.univ-lyon1.fr/} database and {\it
  NASA Extragalactic Database
  (NED)}\footnote{http://nedwww.ipac.caltech.edu/} for distance
measurements and morphological classifications. For four galaxies we
do not have distance measurements and 10 galaxies are classified as
either elliptical or spiral in both {\it NED} and {\it Hyperleda}
databases. After excluding these galaxies, we are left with a final
sample of 371 lenticular galaxies for which we report our analysis in
this Letter. We have not applied any inclination cut on our sample
galaxies.  The sample, while not complete, is representative of
lenticular galaxies in the nearby universe and with the availability
of multi-wavelength data is an unprecedented resource to study
lenticular galaxy properties.

We retrieved the data for all sample galaxies in the form of images
and photometric/spectroscopic measurements from SDSS and 2MASS data
archives. The magnitudes reported here are not corrected for galactic
extinction and K-correction (which will be small because all
galaxies have $z \le 0.05$). In Figure~\ref{f1} we show the
distribution of total absolute magnitude ($M_K$) in the $K_s$ band for
the sample, which is seen to span a wide range in luminosity ($-26.5 <
M_K < -17.5$). We divide the sample into faint and bright groups,
using $M_K = -24.5$ as a boundary following Barway \etal
(2007;2009). The boundary at $M_K = -24.5$ is somewhat arbitrary but our
results do not critically depend on small ($\sim 0.5$ mag) shifts in
the dividing luminosity. The bright group has 160 (43 \%) lenticular
galaxies while the remaining 211 (57 \%) lenticular galaxies belong to the faint
group. Using a typical colour of
$r-K = 3.0$ for early-type galaxies (Fukugita \etal 1995; McIntosh
\etal 2006) this luminosity division in $K$ band corresponds to $M_r =
-21.5$ in SDSS $r$ band. Using this, we divide the sample into faint
and bright groups in the optical as well. According to this luminosity
division the bright group has 163 (44 \%) lenticular galaxies while
208 (56 \%) lenticular galaxies belong to the faint group
(Figure~\ref{f2}), with absolute $r$ magnitudes being in the range
$-23.5 < M_r < -14.5$.

\begin{figure}
\centerline{\psfig{figure=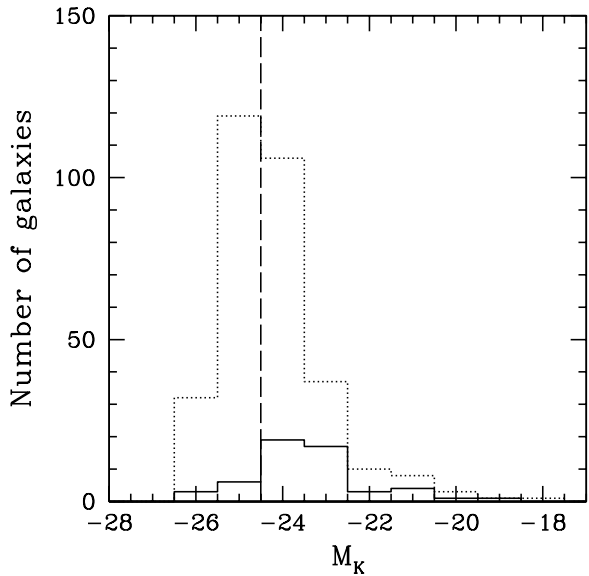,width=8.0cm,angle=0.}}
\caption{Distribution of barred (solid line) and unbarred galaxies
(dotted line) as a function of total absolute magnitude ($M_K$). 
The vertical dashed line corresponds to total absolute magnitude 
$M_K$ = -24.5, which we use to divide low- and high-luminosity 
lenticular galaxies. 54 galaxies are classified as barred in the 
{\it HyperLeda} database.}
\label{f3}
\end{figure}


\begin{figure}
\centerline{\psfig{figure=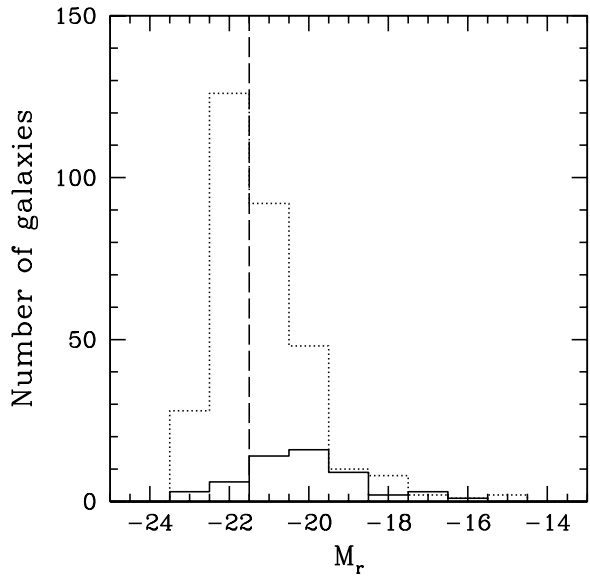,width=8.0cm,angle=0.}}
\caption{Distribution of barred (solid line) and unbarred galaxies 
(dotted line) as a function of total absolute magnitude ($M_r$). The 
vertical dashed line corresponds to total  absolute magnitude 
$M_r$ = -21.5, which we use to divide low- and high-luminosity 
lenticular galaxies. 54 galaxies are classified as barred in the 
{\it HyperLeda} database.}
\label{f4}
\end{figure}


\section{Analysis and results}

Historically, bars were identified by eye, by experts using a variety
of criteria (de Vaucouleurs 1963, Eskridge \etal 2000). The most
widely adopted quantitative technique for identifying bars is the
ellipse-fitting method, in which a bar must exhibit a characteristic
signature in both the ellipticity and position angle profiles
(Marinova \& Jogee 2007; Barazza \etal 2008; Sheth \etal 2008; Knapen
\etal 2000). A simplified version of this technique measures the
difference in the axial ratio and position angles of a best-fit
ellipse to one interior and exterior isophote (Whyte \etal 2002). In
general, the visual and ellipse-fitting methods agree about 85\% of
the time, with egregious disagreement only 5\% of the time
(Menendez-Delmestre \etal 2007, Sheth \etal 2008). In edge-on
galaxies, as Combes \& Sanders (1981) first pointed out, bars result
in boxy- or peanut-shaped bulges (Athanassoula 2005, Bureau \etal
2006). Studying the Fourier modes of the light distribution (Aguerri
\etal 1998, 2000; Laurikainen \etal 2005), or fitting the different
structural components to the surface brightness distribution (Prieto
\etal 2001; Aguerri \etal 2005; Laurikainen \etal 2005; Weinzirl \etal
2009) have also been used to reveal the presence of a bar.

In the present study, we used two methods to identify a galaxy as
barred: (1) we checked if each galaxy was classified as barred in the
{\it HyperLeda} and {\it NED} databases and (2) we independently,
visually, classified each lenticular galaxy as barred or unbarred
using SDSS images in optical and 2MASS images in the near-infrared. In
order to reach a higher S/N than that of the individual images in
different filters, we produced a combined image, as described in
Lisker \etal (2006), by co-adding the $g$, $r$, and $i$ for SDSS and
$J$, $H$, and $K$ for 2MASS by applying weights to each image,
following Kniazev \etal (2004). The visual inspection of these
co-added SDSS and 2MASS images was carried out by us to look for the
presence of a bar. The two classification methods were in close
agreement with a consistency level better than 98\%. A caveat in using
this approach is that our methods do not make any distinction between
strong, weak and nuclear bars. In the few cases of conflict between
our classification and the Hyperleda classification, we use the
Hyperleda one. It must be remembered that bar identification is highly
subjective because (a) one might miss weak bars during visual
classification if the images are too shallow or (b) one might miss
some bars because of inclination effects that we have not
considered. Nevertheless, the consistency between our visual
classification and that in Hyperleda was very good, indicating that
only obvious, strong bars were being identified in both cases .

A bar was found in 15\% of the lenticular galaxies in our sample \ie
54 galaxies were classified as barred in {\it HyperLeda} and {\it NED}
database as well as from our visual inspection of SDSS and 2MASS co-added
images. This should be taken as a lower limit of detection of bars in
lenticular galaxies as the fraction of galaxies classified as barred
strongly depends on the techniques used to detect the bars (Aguerri
\etal 2009). Our detected bar fraction agrees well with a recent study
by Nair \& Abraham (2010) based on a catalogue of detailed visual
classification for 14,034 galaxies in the SDSS DR4. Out of 966
lenticular galaxies in the Nair \& Abraham (2010) sample, 117 (12\%)
were classified as barred which is also consistent with the RC3 visual
strong bar fractions in the local universe. These authors have adopted
the bar classification scheme in which all the bar types are viewed as
definite bars and is more conservative than that of the RC3, where
systems classified as weakly barred include objects that only possibly
contain bars.  The bar fractions reported by these authors as well as
from our study are low compared to some previous studies which quote
bar fractions as high as 60\% (de Vaucouleurs 1963; Aguerri \etal 2009) 
in lenticular galaxies. 

Our analysis finds that the bar fraction in lenticular galaxies
depends on the luminosity. In Figure~\ref{f3} and Figure~\ref{f4} 
we show the distribution of barred and unbarred lenticular galaxies 
as a function of luminosity in 2MASS $K$ and SDSS $r$ band, respectively. 
The distribution for barred lenticular galaxies in both
optical and near-infrared bands reveals that 83\% of the barred
lenticular galaxies belong to the faint group while the bright group
has 17 \% barred lenticular galaxies out of 54 barred lenticular
galaxies in our sample. Bars are found in 21\% of fainter lenticular
galaxies while they are found in only 6\% of more luminous lenticular
galaxies. This suggest that  bars occur preferentially in faint
lenticular galaxies pointing to a possible fundamental difference in
the way in which faint and bright lenticular galaxies are formed
as suggested by Barway \etal (2009;2007).

It would be particularly interesting to know whether environment plays
a role in the dichotomy in bar fraction for the bright and faint
lenticular galaxies that we found in our investigations.  This is
important because lenticular galaxies are more common in high density
environments (\ie groups and clusters) where the influence of
environment greatly affects galaxy disks (Aguerri \etal 2004).  To
investigate this issue, we examine the environment of our sample of
lenticular galaxies and divide our sample into field and group/cluster
environment using data from Tago \etal 2010 which uses the FoF
(friends-of-friends) group search method to search for groups in the
SDSS Data Release 7 (DR7).  Out of 371 lenticulars that we have in our
sample, 108 galaxies are in the field and 263 are members of a group/cluster, 
which reflects the fact that a majority of lenticular galaxies
are located in dense environments. This is true also for the barred
lenticular galaxies in our sample.  Only nine barred lenticular
galaxies are found in field and remaining 45 barred lenticular
galaxies are members of group/cluster.  We do not see a significant
environment dependence for the bright and faint class of our sample
lenticular galaxies. However, barred lenticular galaxies show a
significant environment dependence if one divides galaxies into bright
and faint classes (see Table 1).  For barred lenticular galaxies, our
analysis suggests that faint barred galaxies occur more frequently in
group/cluster environments than their brighter counterparts. This is
well supported by the fact that there is no environment bias for
bright and faint class of our lenticular sample as the fraction of
galaxies are same in field and group/cluster environments is about the
same in both classes. From Table 1 it is clear that for the bright class, 
we have 44\% galaxies in the field and group/cluster. This is also true 
for faint class where a comparable 56\% galaxies are in the field and 
group/cluster.

\begin{table}
\begin{center}
\begin{minipage}{125mm}
\caption{Environment dependence for our sample.} 
\label{table2}
\begin{tabular}{llrr}
\hline
Galaxy Type &               & All S0's    & Barred S0's \\
\hline 
Bright      &  field      & 48/108 (44\%) & 04/09 (44\%) \\
            &  group/cluster & 115/263 (44\%)& 08/45 (18\%) \\
Faint       &  field      & 60/108 (56\%) & 05/09  (56\%) \\
            &  group/cluster & 148/263 (56\%)& 37/45 (82\%) \\
\hline 
\hline
\end{tabular}

Notes.\ group/cluster membership determined by  Tago \etal (2010). 
\end{minipage}
\end{center}
\end{table}

\section{Discussion and Conclusion}
Many observations of disk galaxies, combined with results of simulations, strongly 
suggest that the rearrangement of disk mass into rings and bars funnels gas and 
stars to the centre of the galaxy which is an important driver for the secular evolution 
process (see Kormendy \& Kennicutt 2004; Athanassoula 2005 and references therein for
reviews). Recent studies of the distribution of bar strength has shown that lenticular 
galaxies on average have weaker bars than spiral galaxies in general, and even weaker than 
early-type spirals (Knapen 2010). Barazza \etal (2009) found evidence that the bar fraction 
is related to the morphological structure of the host galaxies in the sense that the bar 
fraction rises from early- to late-type disk galaxies (\ie from bulge-dominated galaxies to 
disk-dominated galaxies) and does not change with redshift. These authors also suggest that 
bars are typically formed or destroyed during processes in which the morphology of the disk 
is emerging or changing. In other words, bars are not dissolved in, for instance, lenticular 
galaxies, but can be destroyed during the processes in which a disk galaxy is transformed 
into a lenticular.  Gadotti \etal (2003) have proposed, 
from the study of two lenticular galaxies without disks using N-body simulation, 
an alternate scenario in which bars can be 
formed in  lenticular galaxies through the dynamical effects of nonspherical 
halos.

For lenticular galaxies, our investigations suggest that the formation of 
bars is a complex process. It not only depends on the total luminosity of
galaxy but environment of the host galaxy also plays a crucial role in bar 
formation and the question whether internal or external factors are more 
important for bar formation and evolution are not easy to answer definitively.

Barway \etal 2009 (also see Boselli \& Gavazzi 2006) have suggested
that faint lenticular galaxies in clusters might be the result of ram
pressure stripping of disk galaxies, where fading of the disc causes a
change of morphology.  Their results obtained using only photometric
data are consistent with the spectroscopic results of Barr \etal
(2007), which support the theory that lenticular galaxies are formed
when gas in normal spirals is removed, possibly when well-formed
spirals fall into a cluster. If lenticular galaxies in clusters are
indeed transformed spirals, it is likely that they preserve other
signatures of their earlier existence, and the presence of a bar could be a
natural expectation if disk galaxies are transforming into faint
lenticular galaxies due to an interaction with the cluster medium and
with other galaxies in the cluster.  This is also consistent with a
scenario in which bars are rather stable and long-lived
structures. 

Recent studies have shown that  fainter, bluer and less massive
disk galaxies have higher bar fractions (Barazza \etal 2008; Aguerri
\etal 2009). Our study finds that a higher bar fraction in lenticular
galaxies occurs at luminosities $M_r > -21.5$ or $M_K > -24.5$. At
this point, it should be noted that the {\it relative} change in bar
fraction between bright and faint lenticular galaxies (not the
absolute value of the fraction) is the relevant parameter, because bar
detection fraction can vary substantially when different techniques
are used to identify a bar.  At poorer signal-to-noise ratio, \ie for
faint galaxies, it should get more difficult to detect a bar. The fact
that we find a larger fraction of bars in faint galaxies indicates that
the effect is real, and may be even stronger, if the bias introduced
by the poorer signal-to-noise ratio is accounted for.

Our results are supported by the study of Mendez-Abreu \etal (2010) 
which suggests that bars are hosted by galaxies in a tight range of
luminosities ($-22 < M_r < -17$) and mass using measures of the bar
fraction in the Coma cluster, a rich environment,  from HST-ACS 
observations. However, in all above studies lenticular galaxies are 
treated as disk galaxies and no effort has been made to study the 
luminosity and environment dependence of host galaxy on bars for 
disk galaxies and lenticular galaxies separately. Detailed analysis of 
bar properties and correlations between bars and various observed properties 
of lenticular galaxies in optical and in near-infrared will appear in a 
forthcoming paper (Barway \etal \ 2010), where we discuss the results 
in the context of galaxy evolution scenarios within the framework of N-body 
simulations and possible links to the formation of classical bulges and 
pseudobulges in lenticular galaxies.

\section*{acknowledgements}

SB thanks Petri Vaisanen for helpful discussions. We thank an anonymous 
referee for insightful comments that have greatly improved both the content 
and presentation of this Letter. This publication makes use of data products 
from the Two Micron All Sky Survey, which is a joint project of the University 
of Massachusetts and the Infrared Processing and Analysis Center/California 
Institute of Technology, funded by the National Aeronautics and Space 
Administration and the National Science Foundation. Funding for the SDSS 
and SDSS-II has been provided by the Alfred P. Sloan Foundation, the Participating
Institutions, the National Science Foundation, the U.S. Department of
Energy, the National Aeronautics and Space Administration, the
Japanese Monbukagakusho, the Max Planck Society, and the Higher
Education Funding Council for England. The SDSS Web Site is
http://www.sdss.org/. The SDSS is managed by the Astrophysical
Research Consortium for the Participating Institutions. This research
has made use of the NASA/IPAC Extragalactic Database (NED) which is
operated by the Jet Propulsion Laboratory, California Institute of
Technology, under contract with the National Aeronautics and Space
Administration. We also acknowledge use of the HyperLeda database.

\end{document}